\documentclass[article,report,preprint,review,prb,letterpaper,amsfonts,amssymb,amsmath,showpacs]{elsarticle}
\usepackage{graphicx}
\usepackage[dvipsnames]{color}
\usepackage{multicol}
\begin{document}

\title{Pressure induced ferroelastic phase transition in LuLiF$_{4}$ compound}
\author[kfu]{Anastasia Petrova}
\address[kfu]{Institute of Physics, Kazan Federal University, Kremlevskaya St.~16a, 420008 Kazan, Russia}
\ead{npetrova131@gmail.com}
\author[kfu]{Oleg Nedopekin}
\address[kfu1]{Centre for Quantum Technologies, Kazan Federal University, Kremlevskaya St.~16a, 420008 Kazan, Russia}
\author[kfu,kfu1]{Dmitrii Tayurskii}
\author[ismans]{Benoit Minisini}
\address[ismans]{Institut Sup\'erieur des Mat\'eriaux et M\'ecaniques Avanc\'es du Mans, 44 Avenue Bartholdi, 72000 Le Mans, France}
\date{\today}

\begin{abstract}
The behavior of LuLiF$_{4}$ sheelite (I4$_{1}$/a, Z = 4) under hydrostatic pressure was investigated by means of the first principles calculations. The ferroelastic phase transition from the tetragonal structure of LuLiF$_{4}$ to fergusonite structure (C$_{12}$/c1, Z = 4) has been found at 10.5 GPa. It has been determined that this is the second order phase transition.
\end{abstract}

\begin {keyword}
A. Fluorides \sep D. Pressure induced phase transitions
\end {keyword}
\maketitle

\section{Introduction}

The interest to fluorite rare-earth compounds (with the scheelite CaWO$_{4}$ structure) increases significantly due to their possible application in laser technologies and microelectronics. The intrinsic dipole moment materials invoke the particular curiosity in a certain temperature range. One of these compounds (namely, LuLiF$_{4}$) has been recently investigated at high pressures by synchrotron angle-dispersive x-ray powder diffraction in a diamond anvil cell at room temperature \cite{Grzechnik2005}. A tricritical phase transition to the fergusonite crystal structure was found at 10.7 GPa, but the type of this phase transition remained unknown. The present article is devoted to the searching for the phase transition in LuLiF$_{4}$ at high pressures by means of density functional theory (DFT) \cite{PhysRev.136.B864, PhysRev.140.A1133} and to identification of its type.

\section{\label{secCalculation parameters}Calculations}

Two possible crystal structures of LuLiF$_{4}$ with symmetries I4$_{1}$/a and C12/c1 were investigated by means of ab-initio calculations. VASP 5.2 (Vienna Ab-Initio Simulation Package)\cite{PhysRevB.54.11169} software package, the part of MedeA\footnote{Materials Design, S.A.R.L.} modeling interface, was used to perform the first principles DFT calculations.
The geometry of structures was optimized in the range of the pressures from 0 GPa to 20 GPa with 2 GPa step until the maximum force dropped below 0.005 eV/{\AA}, whereas the self-consistent field energy convergence criterion was set at ~$10^{-6}$\,eV.

All calculations were performed in ``non magnetical'' mode (i.e. two electrons in each state). The electronic degrees of freedom were described using the projector augmented wave method \cite{PhysRevB.50.17953} and basis of plane waves as implemented in VASP 5.2. The valence electrons of Lu were considered as the ``kept frozen in the core''. The exchange-correlation functional has been approximated by the gradient corrected form proposed by Perdew-Burke-Ernzerhof \cite{PhysRevB.54.16533}. The Dudarev approach \cite{PhysRevB.57.1505} was applied within simplified generalized gradient approximation (GGA)+U scheme \cite{PhysRevB.54.16533}.
The other calculation parameters were chosen the same as in the paper \cite{petrova2012ab}.


\section{\label{secResults}Results}
\begin{figure}[!h]
\begin{center}
\includegraphics[width=120 mm]{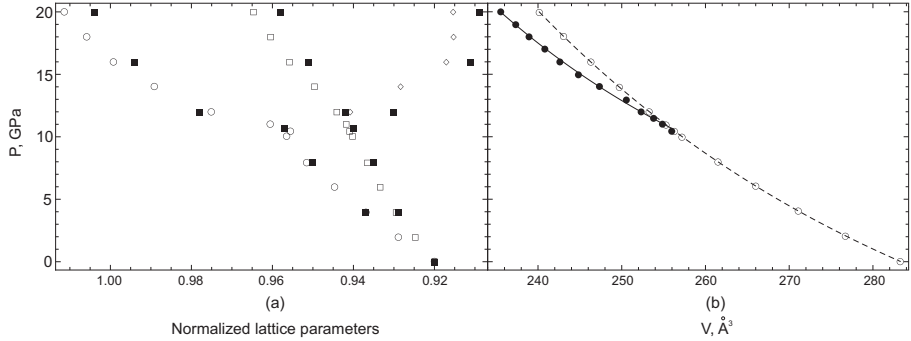}
\caption{\label{fig1} (a) pressure dependencies of the lattice parameters (LuLiF$_{4}$ structure, C12/C1 symmetry) normalized to the respective value at ambient pressure: experimental values \cite{Grzechnik2005}- solid square; ab-initio calculations- empty diamond- $a^m/a_0$, empty square- ~$b^m/b_0$, empty circle- $c^m/c_0$); (b)
the empty circles- the unit-cell volume of I4$_1$/a LuLiF$_{4}$ structure versus the pressure,
the solid circles- the unit-cell volume of C12/c1 LuLiF$_{4}$ structure versus the pressure, dashed and solid lines- Birch – Murnaghan approximation.}
\end{center}
\end{figure}

The analysis of lattice parameters, unit cell volume, order parameter and bulk modulus under pressure has been performed in order to determine the type of phase transition and to obtain the transition pressure. In FIG.~\ref{fig1} (a) the pressure dependencies of the lattice parameters (where $a_{m}$, $b_{m}$ and $c_{m}$ are the lattice parameters of monoclinic structure; a$_0$, b$_0$, c$_0$ are respective values of the lattice parameters at ambient pressure) reflecting the transformation of the structure from I4$_1$/a to the C12/c1 symmetry are shown. Since C12/c1 symmetry group is a subgroup of I4$_1$/a, the structural parameters of these symmetries coincide below 10.5~\, GPa. Based on these results we can conclude that the phase transition occurs at 10.5 GPa that is in a good agreement with the experimental data \cite{Grzechnik2005}.

In order to identify the type of phase transition the volumes of LuLiF$_4$ unit cell for two symmetries were calculated for pressure up to 20 GPa.
FIG.~\ref{fig1} (b) demonstrates the difference of the cell volumes between two symmetries above 10.5~\, GPa. It is well known that the energy and the volume of the system are changing smoothly at the second order phase transition \cite{Landau1936}, which is accompanied by the changes of the system symmetry as in the present case.
Based on FIG.1 we can conclude that the second order phase transition occurs in the analyzed compound LuLiF$_{4}$ at high pressures.

\begin{figure}[h]
\begin{center}
\includegraphics[width=80 mm]{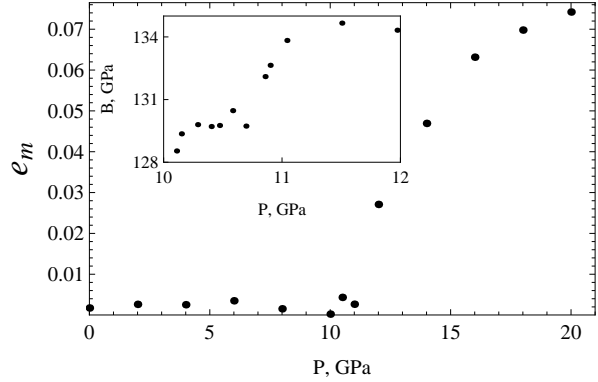}
\caption{\label{fig2} The order parameter of LuLiF$_4$ structure versus the pressure. Inset shows the pressure dependence of the bulk modulus of LuLiF$_4$ structure.}
\end{center}
\end{figure}

The order parameter versus the pressure was plotted to find critical pressure of the phase transition.
The second rank strain tensor components have been selected as a primary order parameter \cite{Grzechnik2005}. The tensor function has the following form $\textbf{e}_m=1/\sqrt{2}(\textbf{e}_{xx}-\textbf{e}_{yy})$ \cite{Tsunekawa1993}. 
The spontaneous strains $\textbf{e}_{xx}$ and $\textbf{e}_{yy}$, contributing to the order parameter $\textbf{e}_m$, are determined as follows: $\textbf{e}_{xx}=(c_m-a_t)/a_t,  \textbf{e}_{yy}=(a_m/\sqrt{2}-a_t)/a_t$ (where $a_t$ is a lattice parameter of tetragonal structure, $a_{m}$ and $c_{m}$ are the lattice parameters of monoclinic structure). The order parameter starts to change smoothly from zero (I4$_{1}$/a symmetry) to a nonzero (C12/c1 symmetry) value at the point~10.5\,GPa (see FIG.~\ref{fig2}).
\



The dependence of bulk modulus B on pressure from 0 GPa to 12 GPa was also investigated. The change in the behavior was observed from 10.5 GPa to 11 GPa on FIG.~\ref{fig2}. The bulk modulus $B=16$ GPa, volume $V=323.4$ ${\AA}^3$, and first pressure derivatives of the bulk modulus $B_0^\prime=9.9$ for C12/c1 symmetry were calculated by fitting the pressure-volume compression data with Birch - Murnaghan equation of state ~\cite{murnaghan1944compressibility}. The same parameters for I4$_{1}$/a symmetry at ambient pressure have been obtained previously \cite{petrova2012ab}. The pressure derivatives of the bulk modulus of two symmetries LuLiF$_{4}$ structure are extremely different.

Analogous compound YLiF$_{4}$ was also studied by means of the DFT method in the previous paper \cite{0953-8984-18-8-009}. It has been shown that YLiF$_{4}$ undergoes the phase transitions from the scheelite phase (I4$_{1}$/a, Z = 4) to the fergusonite-like phase (I$_{2}$/a, Z = 4) and to the LaTaO$_{4}$ like-phase (P2$_{1}$/c, Z~\, = 4). To verify such type of phase transition to the C2/c symmetry in LuLiF$_4$ compound the dependence of the enthalpy on pressure was calculated. The dependence demonstrates the growth of enthalpies of I4$_{1}$/a and C2/c structure symmetries with the pressure. The enthalpy of structure with C2/c symmetry is higher than enthalpy of structure with symmetry I4$_{1}$/a by 0.82 eV per elementary cell. It can be concluded that such type of phase transition is not carried out in LuLiF${_4}$ compound. Also the enthalpy of P12/c1 structure symmetry has been calculated. It turned out to be higher than the enthalpy of structure with symmetry I4$_{1}$/a by 0.76 eV per elementary cell. Thus LuLiF$_4$ structure with symmetry I4$_{1}$/a is energetically most favorable.

\section{Conclusion}
Thus, in the present work the ferroelastic phase transition of LuLiF$_4$ scheelite (I4$_1$/a, Z = 4) structure under pressure was found at 10.5 GPa by means of DFT. The ferroelastic phase transition from the tetragonal structure to fergusonite one (C12/c1, Z = 4) was identified on the base of pressure dependence of the structural parameters, the order parameter and the cell volume. It was determined that this is the second order phase transition. The absence of phase transition to C2/c and to P12/c1 structure symmetries was shown.

\section*{Acknowledgements}
{This work was supported in part by the Ministry of Education and Science of Russian Federation.}

\section*{References}
\bibliography{LiLuF4}
\end{document}